\documentclass[fleqn,usenatbib]{mnras}

\usepackage{newtxtext,newtxmath}

\usepackage[T1]{fontenc}

\DeclareRobustCommand{\VAN}[3]{#2}
\let\VANthebibliography\thebibliography
\def\thebibliography{\DeclareRobustCommand{\VAN}[3]{##3}\VANthebibliography}

\usepackage{graphicx}	
\usepackage{amsmath}	
\usepackage{siunitx}
\usepackage{tabularx,ragged2e,booktabs,caption}
\usepackage{soul} 

\title[Constraining boson\&fermion dark matter stars]{Constraining exotic compact stars composed of bosonic and fermionic dark matter with Gravitational Wave events}

\author[Stephan Wystub et al.]{
Stephan Wystub,$^{1}$\thanks{E-mail: wystub@astro.uni-frankfurt.de}
Yannick Dengler,$^{2}$\thanks{E-mail: yannick.dengler@uni-graz.at}
Jan-Erik Christian$^{1}$\thanks{E-mail: christian@astro.uni-frankfurt.de}
and J\"urgen Schaffner-Bielich$^{1}$\thanks{E-mail: schaffner@astro.uni-frankfurt.de}
\\
$^{1}$Institut f\"ur Theoretische Physik, Goethe Universit\"at Frankfurt, Max von Laue Stra\ss e 1, D-60438 Frankfurt, Germany\\
$^{2}$Institut f\"ur Physik, Universit\"at Graz, Universit\"atsplatz 3, 8010 Graz, Austria\\
}

\date{Accepted XXX. Received YYY; in original form ZZZ}

\pubyear{2022}

\begin{document}
\label{firstpage}
\pagerange{\pageref{firstpage}--\pageref{lastpage}}
\maketitle

\begin{abstract}
We investigate neutron star-black hole (NS-BH) merger candidates as a test for compact exotic objects. Using the events  GW190814, GW200105 and GW200115 measured by the LIGO-Virgo {collaboration}, which represent a broad profile of the masses in the NS mass spectrum, we demonstrate the constraining power for the parameter spaces of compact stars consisting of dark matter for future measurements. We consider three possible cases of dark matter stars: self-interacting, purely bosonic or fermionic dark matter stars, stars consisting of a mixture of interacting bosonic and fermionic matter, as well as the limiting case of {self-bound} stars. We find that the scale of those hypothetical objects are dominated by the one of the strong interaction. The presence of fermionic dark matter requires a dark matter particle of the GeV mass scale, while the bosonic dark matter particle mass can be arbitrarily large or small. In the limiting case of a {self-bound} {constant speed of sound parametrization}, we find that the vacuum energy of those configurations has to be similar to the one of QCD.
\end{abstract}





\section{Introduction}\label{incanus}
Ever since the first direct observation by the LIGO and Virgo collaborations, gravitational waves (GWs) have opened a new opportunity to further the research of astronomical objects and the universe. While the first observation GW150914 \citep{LIGOScientific:2016aoc} not only provided direct proof of the last remaining prediction of Albert Einsteins theory of general relativity, the event also offered deep insight into compact objects and their merging process. For GW150914, the generally accepted explanation of the nature of the compact objects were black holes, as the calculation of the masses, final orbital separation and velocities confidently ruled out the possibility of a merger of neutron stars \citep{LIGOScientific:2016aoc}. However, more exotic, yet so far undetected and thus unverified compact objects, for example boson stars, could not be excluded by the detection \citep{LIGOScientific:2016vlm}, especially since no electromagnetic counterpart was found to coincide with the merger.
This was in stark contrast to the event GW170817 \citep{LIGOScientific:2017vwq}, which was accompanied by a series of multi-messenger events spanning the electromagnetic spectrum and also non-electromagnetic messengers, such as high-energy cosmic rays\citep{LIGOScientific:2017ync}. From this it was concluded that the most likely explanation for the merging compact objects was a system of binary neutron stars, which merged into a hypermassive neutron star remnant that collapsed within {$\approx$ 300 miliseconds} \citep{Margalit:2017dij} into a black hole. This event has provided deep insight into the inner structure of neutron stars and the equation of state (EOS) describing them. {A second event also theorized to be a NS-NS merger, GW190425, was observed in 2019, though it could not be ruled out that one or both of the binary components were black holes} \citep{LIGOScientific:2020aai}. Another event, similarly exciting and groundbreaking as the former observations, was the GW190814 event where the {LIGO-VIRGO} collaboration \citep{LIGOScientific:2020aai} reported a compact binary coalescence involving a $22.2-24.3\,\mathrm{M_\odot}$ black hole and a compact object with a mass of $2.50-2.67\,\mathrm{M_\odot}$ at 90\% confidence level. This places this compact object in the mass gap between the currently known lightest black holes and the heaviest neutron stars with a maximum mass of about 2 $M_{\odot}$\citep{NANOGrav:2019jur,Fonseca:2016tux,Antoniadis:2013pzd,Demorest:2010bx,Nieder:2020yqy,Romani:2021xmb,Romani:2022jhd}. The source was localized at a distance of $241^{+41}_{-45}$ Mpc, which is considered as one possibility to explain the absence of an electromagnetic counterpart for the spin parameters of the this event{, though also the mass ratio of the binary components and the spin parameter of the resulting black hole have a strong effect on a dynamical ejecta}. 

There has been much debate about the nature of this compact object, and some have argued that the secondary was most likely a black hole \citep{Tsokaros:2020hli,Godzieba:2020tjn,Tan:2020ics}. For example, \citep{Fattoyev:2020cws} have shown that a stiff equation of state supporting such a super-massive neutron star is inconsistent with constraints from heavy-ion collisions or from the low deformability of medium-mass stars. \cite{Sedrakian:2020kbi} examined static and fast rotating configurations of hypernuclear stars, yet found that the resulting maximal masses are well below the lower bound of the mass of the secondary in GW190814, \citep{Tews:2020ylw} determined that if GW190814 was a BH-NS merger, it would impose strong constraints on the typical radius of neutron stars, while the assumption of a BH-BH merger is consistent with current knowledge of the neutron star EOS. Additionally, \citep{Nathanail:2021tay} conducted a statistical analysis of GW events and astronomical observations, showing that all measurements are in agreement if the maximum neutron star mass $M_{TOV} = 2.2^{+0.12}_{-0.13}  \,\mathrm{M_\odot}$, while $M_{TOV} > 2.4 \,\mathrm{M_\odot}$ leads to an underproduction of ejected mass.

However, \citep{LIGOScientific:2020aai} did not rule out the possibility of the secondary {component} to be an exotic compact object, for example a gravastar, quark star or boson star. This has lead to several hypothesis being proposed to explain the secondary object. For example, \citep{Most:2020bba} have shown that rapidly rotating neutron stars are in fact able to explain the signal, even without collapsing into a black hole shortly before the merger.
Similarly,\citep{Dexheimer:2020rlp} considered the possibility of a massive rapidly-rotating neutron star with exotic degrees of freedom, such as quarks and hyperons,  while  \citep{Bombaci:2020vgw} have shown that quark stars can reach masses comparable to those of GW190814. {Likewise, \citep{Ivanytskyi:2022oxv} were able to show that  such massive stars can also be formed by an early deconfinement phase transition, which does not require a star to be rapidly rotating.} Additionally, \citep{Zhang:2020zsc} have concluded that the secondary {component} could have been a super-fast spinning pulsar.
Other works have assumed modified gravity, where the modified {Tolman-Oppenheimer-Volkoff (TOV) equations} give rise to higher neutron star masses than possible in general relativity. Examples of  this are Scalar-Vector-Tensor-Gravity as suggested by Moffat \citep{Moffat:2020jic} or $f\left(R\right)$-gravity \citep{Astashenok:2021peo}. However, the measurements of the LIGO and Virgo collaborations have already put a strong bound on the parameter spaces of modified gravity and the possible upper mass of the graviton \citep{Zakharov:2017svs, deRham:2016nuf}. 

In 2021, the LIGO-Virgo collaboration reported two new NS-BH merger events, GW200105 and GW200115 \citep{LIGOScientific:2021qlt}, again without an electromagnetic counterpart, but at $1.9\mathrm{M_\odot}$ and $1.5\mathrm{M_\odot}$ with vastly different secondary masses than in the case of GW190814. We will investigate if the respective secondary compact objects could have been hypothetical stars comprised of dark matter, and use these three events, which cover nearly the entire mass range for neutron stars, to test our models against. 
\cite{Chang:2018bgx} have demonstrated that fermion stars can be formed from primordial density perturbations of dark matter. While they found that not all possible stable solutions to the TOV equation can be formed in realistic scenarios, they nevertheless showed that such exotic objects could {have formed very well and survived until today}. Two new GW events, GW191219 and GW200210  with respective secondary masses of $1.17^{+0.07}_{-0.06}M_\odot$ and $2.83^{+0.47}_{-0.42}M_\odot$ were reported in the Gravitational-Wave Transient Catalog 3 (GWTC-3), collected during the O3 run of LIGO, Virgo and KAGRA {detectors} \citep{LIGOScientific:2021djp}. Those events do lie in a fairly interesting mass range, however, as we will discuss later, the constraining power of those signals is low due to their primary companion stars masses.\\

We consider several different scenarios: First, we discuss the case of stars made up entirely of either bosonic or fermionic dark matter. In this scenario, we assume that the pressure and energy density are proportional to the coupling strength $y$ and square of the number density $n$, i.e. $\propto y^2 \cdot n^2$, and analytically determine the limiting cases. {In this case, dark matter self-annihilation is prevented by two conserved quantum numbers, a similar mechanism as in the Dark White Dwarf stars proposed by \citep{Ryan:2022hku}.}
In the second scenario, we assume the stars to be a mix of bosonic and fermionic dark matter, coupled by a point-like interaction. Similar to the first scenario, we recover an energy density and pressure proportional to the number density squared. After solving the Tolman-Oppenheimer-Volkoff (TOV) equation numerically, we determine bounds on the masses of the dark matter particles and coupling strength. We assume this star to be uncharged with respect to a dark charge so that the number densities of bosons and fermions are balancing each other.
In the third scenario, we consider a {self-bound} linear equation of state, and constrain the respective parameter spaces for the necessary vacuum energy.

\section{Self-interacting boson/fermion stars}
It is known that an object exceeding the Roche-limit and breaking apart due to tidal disruption leaves a clearly distinct signal in the wave event, where the amplitude is affected by the number of fragments as $1/N$ \citep{Maggiore:2018sht}. Since such  a signal was not observed during GW190814, GW200105 and GW200115, we can conclude that the respective secondaries did not break apart, in the following we use the Roche-radius as an estimate for the radii of the objects. From the masses of the objects, we can estimate the Roche-radius of the secondary partner by 
\begin{equation}
R = \frac{2GM}{c^2} \left( \frac{2M}{m} \right)^{-1/3},
\end{equation} where $M$ and $m$ are the respective masses of the larger and smaller partner binary objects. This is a non-relativistic expression, and it has been numerically shown that the fully relativistic equivalent to the Roche-limit is spin dependent and considers effects from e.g. the innermost stable circular orbit (ISCO), whereas the critical radius could be above as well as below the ISCO (see \citep{Pannarale:2010vs} and references therein). As a working hypothesis we consider this non-relativistic approximation to investigate how restrictions of the radius constrict the possible parameter spaces of compact objects consisting of exotic dark matter. {Future detections by next generation gravitational wave detectors might enable the observation of the post-merger signal, yielding information to tighten such radius constraints.}

In the case of purely self-interacting fermionic or bosonic dark matter, we consider an energy denstiy of the form 
\begin{equation}
\varepsilon = \varepsilon_{kin} + c\cdot n^2
\end{equation} and a pressure
\begin{equation}
P = p_{kin} + c\cdot n^2
\end{equation} where $n$ is the number density and $c$ is a constant with dimensions $\left(1/\text{mass}\right)^2$. {It should be noted that $p_{kin} = 0$ in the bosonic case.} This yields a noninteracting Fermi gas in the non-relativistic limit, while in the ultra-relativistic limit it approaches the limiting causal EOS $P = \varepsilon$, which has a speed of sound equal to the speed of light, the maximal causal value. For a derivation of this EOS, see Appendix \ref{Appendix}.

For this EOS, analytical expressions for the maximum mass and the critical radius can be found for arbitrary values for the interaction strength $y = \sqrt{c}/m_f$, where $m_f$ is the mass of the fermionic dark matter particle. For stars consisting of self-interacting fermionic dark matter \citep{Narain:2006kx}, these are
\begin{align}
M_{max}^{Fermion} &= \left( 0.384 + 0.165 \cdot y \right) \cdot \frac{m_P^3}{m_f^2} \\
R_{crit}^{Fermion} &= \left( 3.367 + 0.797 \cdot y \right) \cdot \frac{m_P}{m_f^2}
,\end{align}
with $m_P = \sqrt{\hbar c/G}$ the Planck mass.
{The maximum compactness is limited in the case of $y=0$ by $C = GM_{\max}/R_{crit} = 0.114$ and by $C=0.207$ in the case of $y>>1$.} The corresponding expressions for boson stars are

\begin{align}
M_{max}^{Boson} &= 0.164 \cdot y \cdot \frac{m_P^3}{m_b^2}\\
R_{crit}^{Boson} &= 0.763 \cdot y \cdot \frac{m_P}{m_b^2},
\end{align}
as discussed in \citep{Agnihotri:2008zf}, where $m_b$ is the mass of the bosonic dark matter particle and $y = \sqrt{c}/m_b$. Note that the maximum compactness for those boson stars is then $C = 0.215$. As the secondary components in the {LIGO-Virgo} events GW200105 and GW200115 were reported with critical compactness of $C = 0.22$ and $C=0.26$ respectively (see Table \ref{tab:events}), they cannot have been constituted by purely self-interacting fermionic or bosonic dark matter, while the secondary component in the GW190814 event can still be explained by this model as the compactness is only $C=0.14$. {Such bosonic dark matter stars may be formed by gravitational instabilities in the early universe, see e.g. \citep{Khlopov:1985jw}.}\\

\begingroup
\setlength{\tabcolsep}{6pt} 
\renewcommand{\arraystretch}{1.5} 
\begin{table}
\begin{center}

\begin{tabular}{ |c |c| c| c|}
 Event & R [km] & $M_{max} \left[ M_{\odot} \right]$ & Compactness C\\ 
 \hline\hline   
GW190814       & 26  & $2.59^{+0.08}_{-0.09}$ & 0.14  \\
GW200105 & 13 &	 $1.91^{+0.33}_{-0.24}$ & 0.22	\\
GW200115 &   8.5 & $1.5^{+0.85}_{-0.29}$ & 0.26	\\
GW191219 & 24.4 & $1.17^{+0.07}_{-0.06}$  &  0.07\\
GW200210 & 27.6 & $2.83^{+0.47}_{-0.42}$ & 0.15 \\
\end{tabular}
\end{center}
\caption{\protect\footnotesize Calculated Roche Radius R and Maximum Mass M for the events GW190814, GW200105 and G201015}\label{tab:events}
\end{table}

Assuming $M_{max} > 2.5 \mathrm{M_{\odot}}$ and $\mathrm{R_{crit}} < 23 \,\mathrm{{km}}$ for GW190814, this constrains the fermionic interaction strength and mass to a narrow region in the parameter space with a minimum at $m_f = 0.81 \, \mathrm{GeV}$ and  $f = m/y = 0.16\, \mathrm{GeV}$ (see Figure \ref{fig:m-y-exclusion}), which corresponds to $y = m/f = 3.6$. For the purely bosonic case, the interaction strength and mass are constrained to a slightly wider region and can both be arbitrarily small. Therefore, all hypothetical objects which meet the bosonic condition are eligible, while values close to $f_i \approx \sqrt{2} f_\pi$ are required for the fermionic case, where $f_\pi = 92 MeV$ is the pion decay constant.

\begin{figure}
	\begin{center}
	\includegraphics[width=9.cm]{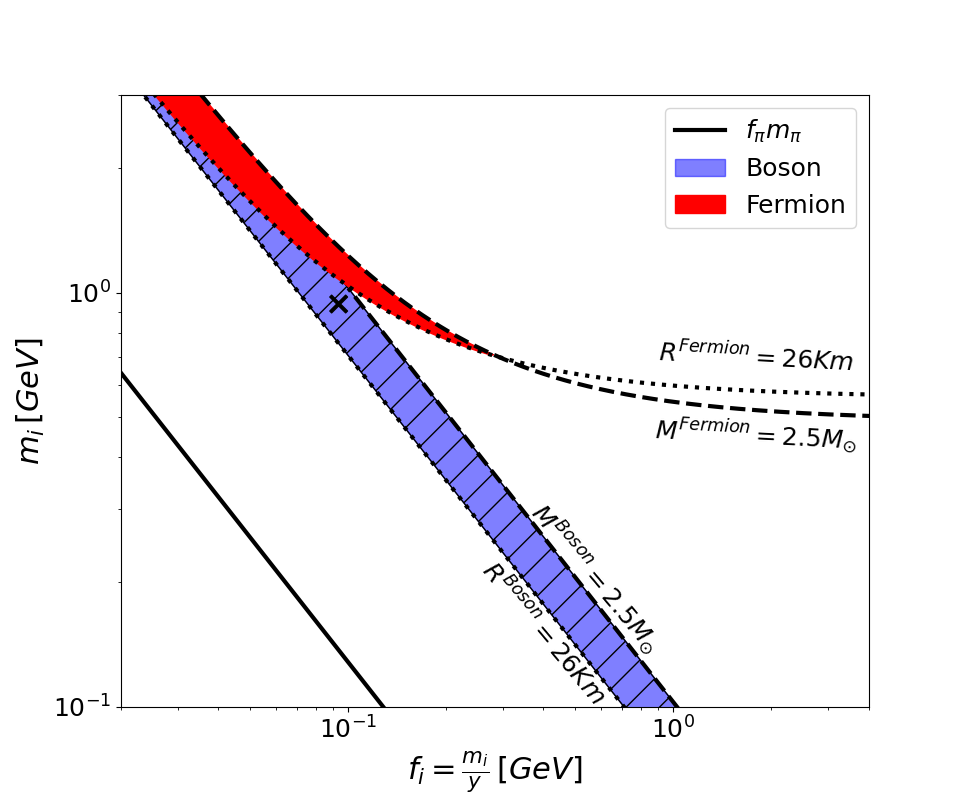}
	\end{center}					
\captionsetup{justification=raggedright}
\caption{\protect\footnotesize Exclusion plot for interaction strength $f = m/y$ and particle mass m assuming $M_{max} > 2.5 \mathrm{M_{\odot}}$ and $\mathrm{R_{crit}} < 26 \,\mathrm{{km}}$. The allowed region for the fermionic case has a minimum at $f = 0.16 \, \mathrm{GeV}$ and $m_f = 0.81 \, \mathrm{GeV}$. The solid black line denotes $f_i m_i = f_{\pi}m_{\pi}$, while the black cross denotes the point where $m_i$ is the nucleon mass and $f_i = f_{\pi}$. 
}
\label{fig:m-y-exclusion}
\end{figure}

From this, it is unlikely that the secondary in GW190814 was an exotic boson star purely consisting  of Peccei-Quinn-Axions \citep{Peccei:1977ur} with a mass of $\approx \SI{1}{keV}$, as this boson mass would imply an interaction strength $y \approx 10^{-11}$ or a decay width $f_a \approx \SI{e5}{GeV}$. Similarly, since the crucial scaling relation for QCD-like axions is $f_i \cdot m_i \approx f_\pi \cdot m_\pi$ \citep{Zyla:2020zbs,DiLuzio:2020wdo}, stars consisting of these axions would not be compatible within our working hypothesis with the measured constraints for radius and mass.
It is also interesting to note that in the case of neutrons comprising fermionic matter, the interaction strength $f$ is required to lie in the range of $\left[ \sqrt{2}f_{\pi}, 2f_{\pi} \right]$, of the order of interaction strengths in neutron star matter. 

For the event GW191219, even though the secondary mass is in a very interesting range, the constraining power is low, since the mass of its primary {object} results in a high estimate for the Roche-radius, thus leaving a large range of possible parameters. For GW200210, both maximum mass and Roche-radius are very similar to GW190814, not offering any more insight into constraining the parameter spaces.

\section{Compact stars consisting of boson-fermion dark matter}
We now consider uncharged objects consisting of a mixture of bosonic and fermionic dark matter. Similar configurations are realized in nature in the form of white dwarfs, where the nuclei are comprising the role of the bosonic matter, while electrons are fulfilling the part of fermionic matter, since the contribution of the lattice of nuclei in the interior has a negligible effect on the EOS. Another example of such a configuration are pion stars proposed by \citep{Brandt:2018bwq}, where a charged condensate of pions and a gas of charged leptons and neutrinos fulfill the roles of bosonic and fermionic matter, respectively. We will assume those dark matter stars to be globally uncharged stars while the number of fermions and bosons are equal, which requires the existence of a dual electromagnetism mediating between the two particles via dark photons. One such proposed and well motivated mechanism is the one of \textit{mirror dark matter} \citep{Berezhiani:1995am,Chacko:2005pe}, for example.\\

Similarly to section 2, we assume a dimensionless energy density and pressure proportional to the square of the interaction strength and number density:
\begin{equation}
\begin{array}{ll}
\varepsilon =& \, \varepsilon_{kin}^{boson} + \varepsilon_{kin}^{fermion} + y^2 n_B^2 
\end{array}
\end{equation}

\begin{equation}\label{pressure}
\begin{array}{ll}
P =  & \, p_{kin}^{fermion}+  y^2 n_B^2 {,}
\end{array}
\end{equation}
where $y$ is the interaction strength and $n_B$ is the number density of bosons. Such an equation of state can be motivated using thermodynamic relations (see Appendix A),  resulting in an EOS that can be rewritten in terms of the squares of the number densities of fermions and bosons. The assumption of charge-neutrality requires the number densities of fermionic and bosonic dark matter to be equal, i.e. $n_F = n_B$. 

Solving the Tolman-Oppenheimer-Volkoff equation and assuming $M_{max} > 2.6 \mathrm{M_{\odot}}$ and $\mathrm{R_{crit}} < 26 \,\mathrm{{km}}$ in the case of GW190814, we find solutions for the coupling strength $f$ and the respective dark matter particle masses (see Figure \ref{fig:MR5}). One can see that the coupling strength $y$ is constrained to a range of $[2.5 \, ,22.5]$. Interestingly, a five-fold increase in $y$ can result in a change of two orders of magnitudes in the respective masses. Also, one of the two dark matter particle masses is always of the order of the GeV-scale, regardless of the coupling strength. This is in stark contrast to the findings in section 2, where the bosonic masses can be arbitrarily small.

In the case of GW200105 and its high compactness, a similar behaviour can be seen (see Figure \ref{fig:MR6}), however, the allowed parameter space is much more restricted compared to GW190814. For masses of the bosonic particle to be smaller than $10^4$ MeV, the interaction strength is constrained to values in the range of $y \approx [10 \, , 20]$. In the case of the even more compact GW200115 secondary {component}, no possible solutions in the parameter space can be found.

\begin{figure}
	\begin{center}
	\includegraphics[width=8.6cm]{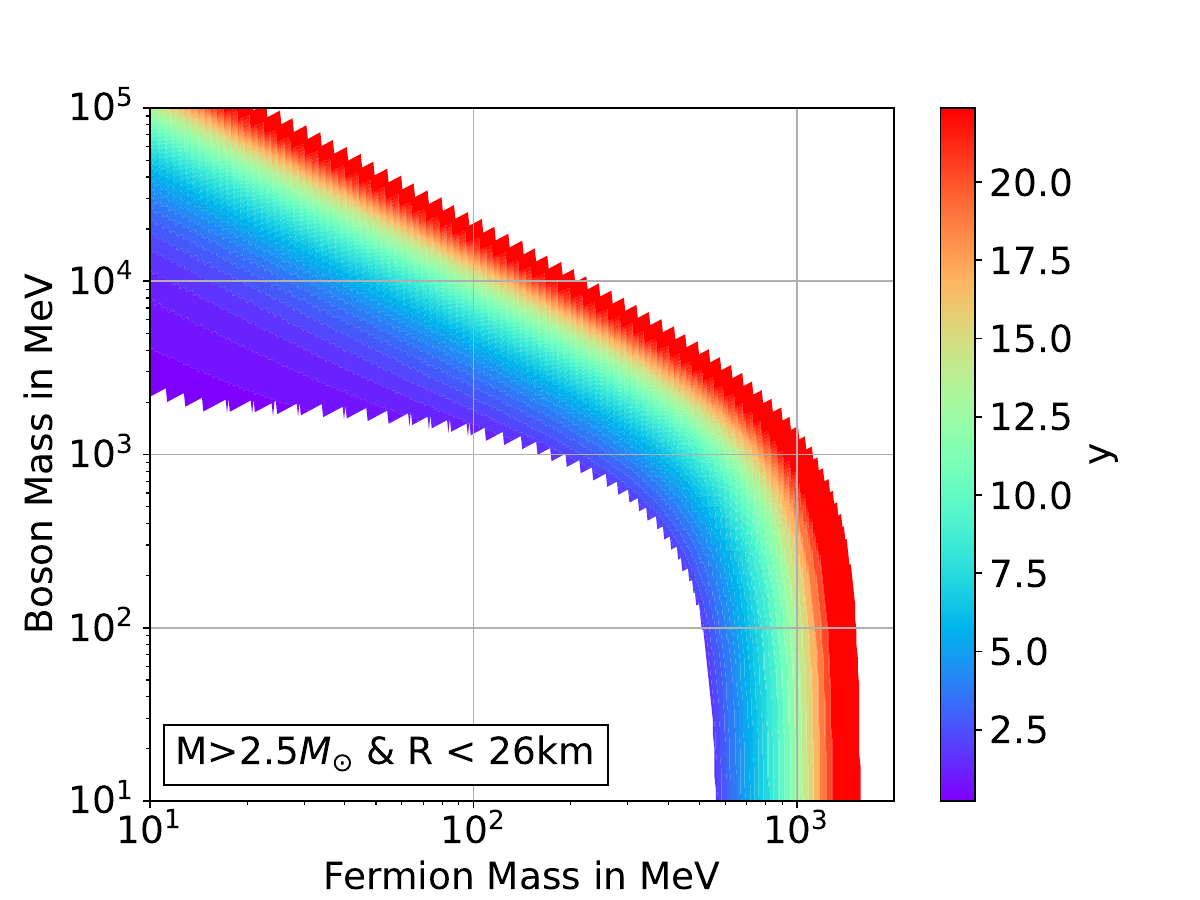}
	\end{center}			

	\captionsetup{justification=raggedright}
	\caption{\protect\footnotesize Exclusion plot for interaction strength y and particle masses $m_f$ and $m_b$ assuming $M_{max} > 2.5 \mathrm{M_{\odot}}$ and $\mathrm{R_{crit}} < 26 \,\mathrm{{km}}$ as in the case of GW190814. As one can see, the region permitting solutions always requires one of the particle masses to be of order GeV. Note that a change of {one order} of magnitude in y results in only small changes in allowed regions for particle masses. }

\label{fig:MR5}
\end{figure}

\begin{figure}
	\begin{center}
	\includegraphics[width=8.6cm]{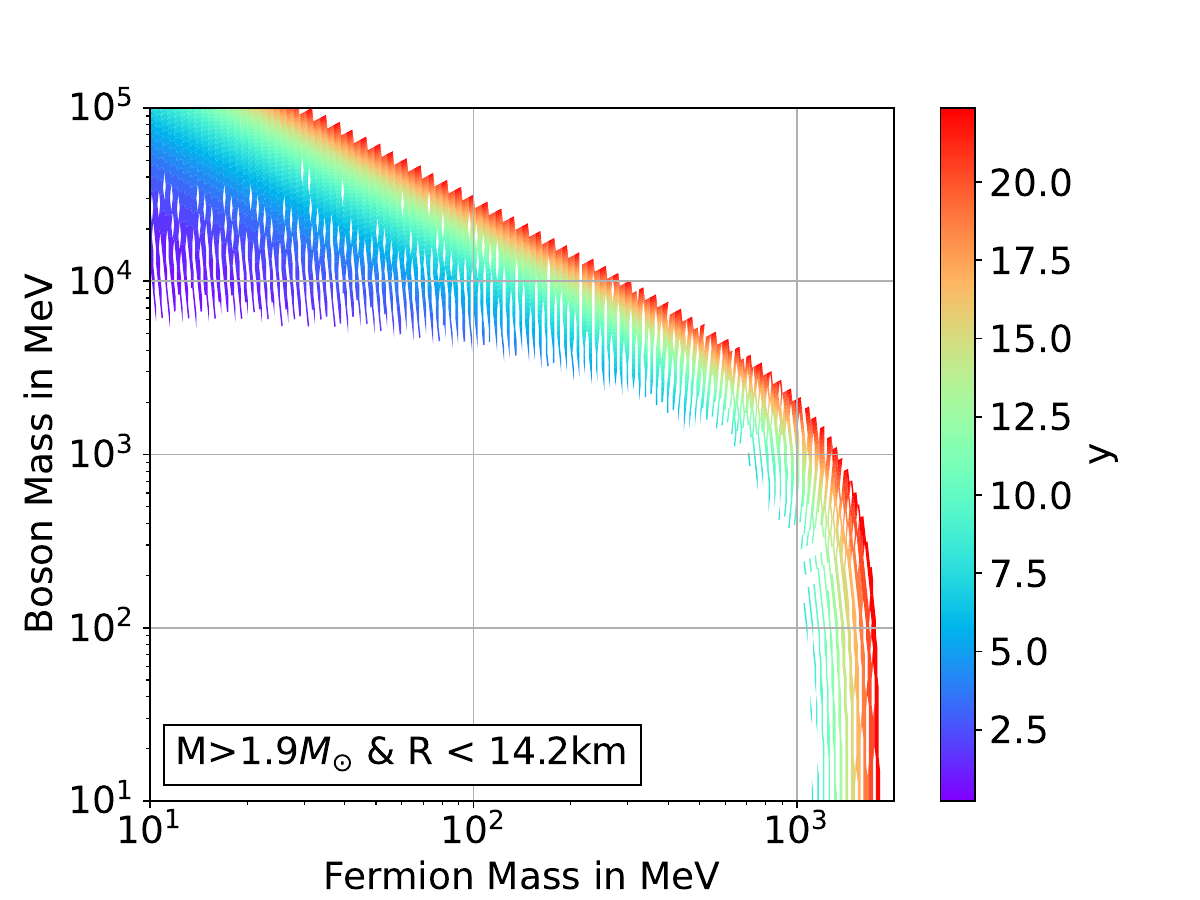}
	\end{center}	
				
	\captionsetup{justification=raggedright}
	\caption{\protect\footnotesize Exclusion plot for GW200105 with interaction strength y and particle masses $m_f$ and $m_b$ assuming $M_{max} > 1.9 \mathrm{M_{\odot}}$ and $\mathrm{R_{crit}} < 13 \,\mathrm{{km}}$. Similarly to GW190814, a change of {one order} of magnitude in y results in only small changes in allowed regions for particle masses. In the case of GW200115 with $M_{max} > 1.5 \mathrm{M_{\odot}}$ and $\mathrm{R_{crit}} < 8.5 \,\mathrm{{km}}$, no valid solutions are found.}

\label{fig:MR6}
\end{figure}

\section{Self-bound Stars}
For a {self-bound} linear equation of state of the form $p = s \cdot \left( \varepsilon - \varepsilon_0 \right)$, the maximum mass and critical radius depend on  the speed of sound $s = c_s^2 = \frac{\partial p}{\partial \varepsilon}$. This equation of state is related to the {MIT bag model} with a bag-constant $\varepsilon_0 = 4B$ and $s=1/3$. Note that the defining feature of a {self-bound} star is that the pressure vanishes at non-vanishing energy density. With this EOS, one finds \citep{Narain:2006kx,Agnihotri:2008zf}

\begin{align}
M_{max} = 2.57 \, M_{\odot} \cdot \left( \frac{\varepsilon_{nm}}{\varepsilon_0} \right)^{1/2} &  \quad s = 1/3 \\
M_{max} = 4.23 \, M_{\odot} \cdot \left( \frac{\varepsilon_{nm}}{\varepsilon_0} \right)^{1/2} & \quad  s = 1 
\end{align}
where $\varepsilon_{nm} = 140\, \textrm{MeV} \, \textrm{fm}^{-3} $ is the energy density of saturated nuclear matter. Given a maximum mass, these expressions analytically provide an upper bound for $\varepsilon_0$. In the case of a minimum radius, the TOV-equations have to be solved numerically to constrain $\varepsilon_0$ from below (the results are shown in Tables \ref{tab:title}, \ref{tab:title 2} and Figure \ref{fig:self-bound13}.) \\

\begin{table}

\begin{center}
s = 1/3
\scalebox{.8}{
\begin{tabular}{ |c |c| c| c| c| }
 Event & R [km] & $M_{max} \left[ M_{\odot} \right]$  & min $\varepsilon_0 [\, \textrm{MeV\,} \textrm{fm}^{-3}]$ & max $\varepsilon_0 [\, \textrm{MeV\,} \textrm{fm}^{-3}]$\\ 
 \hline\hline   
 GW190814       & 26  & 2.5      &  34    &  137   \\
 GW200105 &  13 & 1.9	& 174	& 256\\
GW200115 &  8.5 & 1.5	 & 401	& 411\\
\end{tabular}
}
\caption{\protect\footnotesize Roche Radius R, Maximum Mass and constrained intervals of $\varepsilon_0$ for the events GW190814, GW200105 and G201015 in the case of the speed of sound $s = 1/3$}\label{tab:title}
\end{center}

\begin{center}
s = 1
\scalebox{0.8}{
\begin{tabular}{ |c |c| c| c| c| }
 Event & R [km] & $M_{max} \left[ M_{\odot} \right]$ &  min $\varepsilon_0 [\, \textrm{MeV\,} \textrm{fm}^{-3}]$ & max $\varepsilon_0 [\, \textrm{MeV\,} \textrm{fm}^{-3}]$\\ 
 \hline\hline   
GW190814       & 26  & 2.5   &  39    &  370 \\
 GW200105 & 13 &	 1.9	& 213	& 694\\
GW200115 &   8.5 & 1.5	& 576	& 1113\\
\end{tabular}
}
\caption{\protect\footnotesize Roche Radius R, Maximum Mass and constrained intervals of $\varepsilon_0$ for the events GW190814, GW200105 and G201015 in the case of the speed of sound $s =1$}\label{tab:title 2}
\end{center}
\end{table}

In the case of GW190814, {the required values of $\epsilon_0$ to find solutions} are of the order of the standard MIT bag value $\approx 57.5 \, \textrm{MeV\,} \textrm{fm}^{-3}$. As one can see, the necessary $\varepsilon_0$ is in the range of the QCD scale. This suggests that such {an} object would be dominated by {interactions of a physical scale comparable to QCD interactions}, and that weak interactions do not play a significant role in such {self-bound} stars, regardless of their actual microscopic composition. This is however in stark contrast to GW200105 and GW200115, which require higher values for $\varepsilon_0$ to fulfill the imposed conditions. Especially in the case of GW200115 for s=1, the required values for $\varepsilon_0$ can be as high as $1 \, \textrm{GeV\,} \textrm{fm}^{-3}$. This is still in the typical range of QCD energy densities, as e.g. the one of the QCD crossover transition. However, it is not possible to find values for $\varepsilon_0$ that can fit all signals for a given fixed speed of sound. Hence, {self-bound} stars could generally be ruled out in our adopted scenario.

\begin{figure}
	\centering				
\includegraphics[width=8.6cm]{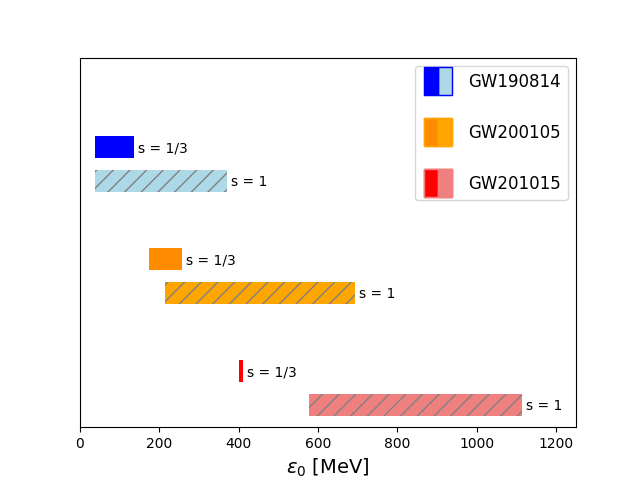}
\caption{\protect\footnotesize Range of allowed values for $\varepsilon_0$ for GW190814, GW200105 and GW200115.}
\label{fig:self-bound13}
\end{figure}

\section{Conclusions}
In this work we investigated NS-BH merger candidates to constrain compact exotic object mergers, by assuming as a working hypothesis that these objects have radii no larger than the Roche-radius. This allows us to demonstrate the constraining power of future radius measurements of merger events for exotic compact objects comprised of dark matter. We studied for this purpose generic models of self-interacting fermion and boson stars. We have shown that in all the studied cases, the characteristic scales necessary to explain GW190814 are very close to the QCD-scale, suggesting that weak interactions do not play a significant role in hypothetical dark matter stars with such high masses and large radii. Additionally, while in the case of self-interacting stars the bosonic dark matter particle mass can be arbitrary, the fermionic dark matter particle mass is required to be in the GeV range. Interestingly, in the case of a mixture of bosonic and fermionic dark matter one of both particle masses is also always in the GeV scale. This implies that if the secondary object in GW190814 was comprised of dark matter with a dark photon mechanism, then the possible dark QCD scale must be similar to the QCD scale. In the limiting case of a {self-bound} linear equation of state, we find again that the vacuum energy needs to be of the scale of QCD.
However, in the case of masses and radii which are common for neutron stars, the applied models are either constrained to very small parameter spaces, or allow for no solutions at all, as demonstrated in the case of the very compact object of GW200115.

\section*{Acknowledgements}

JEC is a recipient of the Carlo and Karin Giersch Scholarship of the Giersch Foundation. SW was supported by GSI Helmholtzzentrum f\"ur Schwerionenforschung Darmstadt under the F\&E program. JEC, JSB and SW acknowledge support by the Deutsche Forschungsgemeinschaft (DFG, German Research Foundation) through the CRC-TR 211 'Strong-interaction matter under extreme conditions' - project number 315477589 - TRR 211. JSB thanks Luciano Rezzolla for helpful discussions.

\section*{Data Availability}

The data underlying this article will be shared on reasonable request to the corresponding author


\newpage



\bibliographystyle{mnras}
\bibliography{Paper} 




\appendix
\section{Derivation of the equation of state}\label{Appendix}
We consider the internal energy $U$ in the microcanonical ensemble given by the fundamental equation of thermodynamics:
\begin{equation}
 \mathrm{d}U = - P \cdot \mathrm{d}V + T\cdot \mathrm{d} S + \sum_i \mu_i \cdot \mathrm{d}N_i,
\end{equation}
where $N_i$ is the particle number of the respective species of particles and $\mu_i$ the corresponding chemical potential.

Now we assume the internal energy density to be quadratic in the number densities of two species of particles, one bosonic and one fermionic:
\begin{equation}
u = u_{kin} + y_1^2 \cdot n_F^2 + y_2^2 \cdot n_B^2,
\end{equation}
with $n_F$ and $n_B$ denoting the number densities $\mathrm{d}N_i/\mathrm{d}V$ of the particles.

The kinetic term for the fermionic particle can be found by integrating the relativistic energy momentum relation over all momenta

\begin{equation}
u_{kin} = \frac{g}{\left( 2\pi \right)^3} \int E\left(\vec{k} \right) \, \mathrm{d}^3\vec{k} = \frac{g}{ 2\pi^2 } \int_{0}^{k_{Fermi}} k^2 \sqrt{{k}^2+m^2} \, \mathrm{d}k,
\end{equation}

where 
\begin{equation}
k_{Fermi} = \left( \frac{6\pi^2}{g} n_F \right)^{1/3}
\end{equation}

is the Fermi-momentum and with $g$ denoting the internal degrees of freedom of the fermionic particle, while for the bosonic particle it only consists of a mass term

\begin{equation}
u_{kin} = m\cdot n_B.
\end{equation}

Thus, we find for the chemical potentials

\begin{align}
& \mu_F = \frac{\partial U}{\partial N_F} = \frac{\partial u}{\partial n_F} =  \frac{\partial u_{kin}}{\partial n_F} + 2\cdot y_1^2 \cdot n_F\\
& \mu_B = \frac{\partial U}{\partial N_B} = \frac{\partial u}{\partial n_B} = \frac{\partial u_{kin}}{\partial n_B} + 2\cdot y_2^2 \cdot n_B.
\end{align}

By making use of the relation $p = \sum_i n_i \cdot \mu_i - u $ we find the total pressure, given as
\begin{equation}
P_{total} = p_{kin} + P = p^{Fermion}_{kin} + p_{kin}^{Boson} + y_1^2 \cdot n_F^2 + y_2^2 \cdot n_B^2.
\end{equation}

Since $p_{kin}^{Boson} = 0$, and setting $y_1^2 + y_2^2 =y^2$ as well as $n_F = n_B$ we arrive at equation \ref{pressure}.
\begin{equation}
P = p^{Fermion}_{kin} + y^2 \cdot n_B^2.
\end{equation}


\bsp	
\label{lastpage}
\end{document}